\documentclass{article}

\usepackage{arxiv}

\usepackage[utf8]{inputenc} 
\usepackage[T1]{fontenc}    
\usepackage{hyperref}       
\usepackage{url}            
\usepackage{booktabs}       
\usepackage{amsfonts}       
\usepackage{nicefrac}       
\usepackage{microtype}      
\usepackage{lipsum}
\usepackage{natbib}
\usepackage{mathptmx}       
\usepackage{helvet}         
\usepackage{courier}        
\usepackage{type1cm}        

\usepackage{makeidx}         
\usepackage{graphicx}        
\usepackage{multicol}        
\usepackage[bottom]{footmisc}

\title{From Digitalization to Data-Driven Decision Making in Container Terminals}

\author{
 Leonard Heilig \\
  Institute of Information Systems\\
  University of Hamburg\\
  Hamburg, Germany \\
  \texttt{leonard.heilig@uni-hamburg.de} \\
   \And
  Robert Stahlbock \\
  Institute of Information Systems\\
  University of Hamburg\\
  Hamburg, Germany \\
  \texttt{robert.stahlbock@uni-hamburg.de} \\
   \And \hspace{3cm}
    Stefan Vo{\ss} \\
  \hspace{3cm}Institute of Information Systems\\
  \hspace{3cm}University of Hamburg\\
  \hspace{3cm}Hamburg, Germany \\
  \hspace{3cm}\texttt{stefan.voss@uni-hamburg.de} \\
   \And
}

\begin{document}

\maketitle
\begin{abstract}
With the new opportunities emerging from the current wave of digitalization, terminal planning and management need to be revisited by taking a data-driven perspective. Business analytics, as a practice of extracting insights from operational data, assists in reducing uncertainties using predictions and helps to identify and understand causes of inefficiencies, disruptions, and anomalies in intra- and inter-organizational terminal operations. Despite the growing complexity of data within and around container terminals, a lack of data-driven approaches in the context of container terminals can be identified. In this chapter, the concept of business analytics for supporting terminal planning and management is introduced. The chapter specifically focuses on data mining approaches and provides a comprehensive overview on applications in container terminals and related research. As such, we aim to establish a data-driven perspective on terminal planning and management, complementing the traditional optimization perspective.
\end{abstract}


\section{Introduction}
\label{sec:1}

In recent decades, terminal operators have strongly invested in automation and digitalization to improve the operational efficiency of their container terminals. While information systems have already become indispensable for terminal planning and management, the current wave of digitalization strives for a better integration and transparency amongst all parts of the supply chain. Extending terminal infrastructure and equipment with sensors, actuators, and mobile technologies, e.g., lead to new levels of transparency allowing to constantly monitor and control resources, cargo flows, and the environment. Existing data, such as from terminal operating systems, together with data from a variety of new data sources, including those available from external systems, however, mostly remain under-processed or under-analyzed to be of real value. 

One central aspect of human intelligence is the ability of learning, i.e., of inferring repeating patterns and relationships from observations that are often reflected in data. Especially in the highly competitive environment of maritime ports and container terminals, it seems that discovering \textit{patterns},  \textit{regularities} or even \textit{irregularities} in operational data has become even more important. Many data-analytic methodologies, approaches and algorithms have been developed with different terminology and objectives. The main common topic among all -- whether an approach is called \textit{data mining} or \textit{machine learning}, etc. -- is to estimate or learn good data-analytic models of real business phenomena in order to provide description, analysis, understanding, prediction, and prescription. This can be useful for decision making in organizations. Traditionally, quantitative research has been focused on the optimization of logistics processes and operations in container terminals using operations research methods \citep[see, e.g.,][]{steenken2004container,stahlbock2008operations}. Given that the amount of complex data is growing, data-analytic models for extracting information and knowledge from data before it is used in optimization routines is set to become a vital factor in improving port operations. At present, however, we only find a few studies presenting data-driven approaches for understanding and addressing problems in container terminals or other port-related operations. Besides mathematical models and optimization methods, business analytics considers the use of data, information technology, statistical analysis, and computer-based data-analytic models to gain improved insights about business operations and to make better decisions.


Note that business analytics is only one part of the even more general digitalization in the maritime industry. Since the beginnings of the containerization in the 1960s, efficient cargo flows rely on efficient information flows. Many port-related information systems have been established to facilitate intra- and inter-organizational information flows in different phases of the container transport. Nowadays, collecting and storing transactional data is not a main challenge, neither from a technical point of view nor as a financial one. However, the challenge is to collect and organize only the relevant data with respect to specific business problems and to derive new knowledge out of it, thus creating value (in combination with required domain knowledge of managers and decision makers). To address current problems in seaports, the maritime industry increasingly recognizes the value of information and respective decision support tools. Thus, digital transformation in ports is nowadays not only focused on collecting more data generated along the logistics chains, but especially on facilitating a clever usage of data in order to gain competitive advantages, such as by adapting business models and improving customer experience, processes, and costs. 

Novel concepts and technology drivers of the current phase of digitalization, such as related to the internet of things (IoT), cloud computing, mobile technologies, and big data, imply huge potentials and challenges in transforming port operations. Those concepts and technologies have not only increased the number of potential data sources and the amount of captured structured and unstructured data (e.g., by using mobile devices and integrating sensors and actuators in port operations), but also provide affordable and highly scalable data storage and processing services (e.g., offered by public cloud providers). It can be observed that costs of storing data are below the costs of deciding which data should be kept and which should be deleted, resulting in a massive flood of data. Some parts are important and valuable, some are not, some not today but maybe in the future. Furthermore, highly scalable computational processing power allows for performing (nearly) real-time analytics at appropriate costs. In this context, the current phase of digital transformation in maritime ports, referred to as the generation of smart procedures \citep{heilig2017analysis}, aims to adapt novel concepts, information technologies, tools, and methods providing means of advanced gathering, processing, and analysis of (real-time) data in order to better understand, plan, control, and coordinate port operations. This transformation shall allow to better utilize port-related resources, equipment, and space, on the one hand, but also an improved information exchange and collaboration within and between maritime ports. 

After giving a brief overview on developments and trends of the digital transformation in ports, this chapter provides an introduction to business analytics and its application in container terminals. With respect to different operations areas of the container terminal, divided into quayside, yard, and landside, we discuss potential data sources and provide an overview of data mining applications that are currently discussed in academia and practice. Thus, the chapter is primarily focused on data mining approaches for descriptive, predictive, and prescriptive analytics in container terminals. Although a lack of studies can be identified, the chapter reviews all relevant works from academic literature addressing terminal-related problems with data mining methods. As such, the chapter represents, to the best of our knowledge, the first state-of-the-art review of data mining applications in container terminals. The remainder of this chapter is structured as follows. Section~\ref{sect:digitalizationports} outlines the main developments of three generations of digital transformation in maritime ports and discusses current trends of the digitalization. The concept and methodology of business analytics is introduced in Section~\ref{sect:businessanalytics}. Section~\ref{sect:dataminingterminal} outlines various data mining approaches with respect to the different terminal operations areas and discusses related literature. Finally, a conclusion and outlook are provided in Section~\ref{sect:conclusion}.

\section{Digitalization in Maritime Ports}
\label{sect:digitalizationports}

Since the beginning of containerization, the digital transformation of port operations has become indispensable for driving innovation and modernization in maritime ports. The ability to share information between involved actors and to track cargo is critical for reducing uncertainties \citep{zhou2007supply}, increasing reliability \citep{panayides2009port}, and improving the coordination in integrated transport processes \citep{lai2008coordination,crainic2009intelligent,Wiegmans2017}. Moreover, advanced information systems can provide a basis for addressing environmental sustainability in martime ports \citep{mansouri2015multi,heilig2017tre}. Due to their important role in achieving a competitive edge, a plethora of information systems and technologies have been adopted in port operations in recent decades. \cite{Heilig2016} provide a comprehensive overview of those port-related solutions. Although past developments have led to a high degree of automation and digitalization, especially in container terminals, there is still considerable potential for improvement. In particular, a better integration of existing information systems and data sources as well as a more intelligent use of data may help to improve planning, controlling, and management of intra- and inter-organizational operations and thus may have a considerable impact on supply chains \citep[for further information and examples, the interested reader is referred to][]{heilig_voss2018}. 

The current impact of digitalization can be observed in many contemporary ports. We define the current phase of digital transformation in ports as the generation of \textit{smart procedures} (see Section~\ref{subsect:thirdgen}). One current trend is related to the concepts of \textit{Industry 4.0} and \textit{Logistics 4.0}, which are strongly related to the development of cyber-physical systems and IoT infrastructures. Here, the focus is to measure, monitor, and control physical processes and objects including their environment by means of automation and connectivity. An example is the \textit{smartPORT logistics} project in the Port of Hamburg (Germany). By collaborating with SAP and T-Systems, the Hamburg Port Authority (HPA) has developed a cloud-based platform to improve traffic flows in the port area. This is supposed to be achieved by enhancing the control of port infrastructure (e.g., movable bridges, traffic lights, parking space) through smart devices and IoT technologies (e.g., sensors and actuators) as well as by managing collected traffic data and real-time communication with port community actors via mobile applications \citep{hpa2017}. Due to a lack of a critical mass of users within the port community, however, the introduction of the mobile application failed \citep{ndr2017}.

Moreover, an emerging impact of \textit{big data} can be identified in the maritime industry, which mainly refers to different technologies and techniques to process and analyze large and complex sets of data that exceed the capacity or capability of conventional methods and systems. The Maritime Port Authority of Singapore (MPA), for instance, is collaborating with IBM to tap big data solutions for improving maritime and port operations, e.g., through a prediction of vessel arrival times and a better detection of movements, authorized activities (e.g., pilotage), and unauthorized activities (e.g., illegal bunkering). Although the term is often misinterpreted or used as buzzword in the industry (e.g., as a substitute for data mining or business intelligence, BI), the growing interest in big data is also reflected in products and services of cargo-handling equipment providers and software vendors. Kalmar, a leading provider of cargo-handling equipment and services, has developed a cloud-based platform to display real-time productivity and operational data as well as maintenance information. Navis, a leading vendor of terminal operating systems (TOS), has launched a Terminal BI Portal to better understand the historical and real-time performance of terminal operations and further aims to use machine learning to gain additional insights from the TOS data. Besides, the increasing interest regarding the general topic of digitalization is reflected by diverse hackathons (e.g., World Port Hackathon) and  competitions (see, e.g., PEMA Student Challenge) encouraging students and scholars to develop new and innovative digital solutions for maritime ports. To better understand the development -- from \textit{paperless procedures} to \textit{smart procedures} --  a brief overview about the main phases of the digital transformation is given in the following \citep{heilig2017analysis}. 


\subsection{First Generation (1980s): Paperless Procedures}

Traditionally, paper-based procedures were established for organizing the information flow, which has been labor intensive, time-consuming, error-prone, and costly. To further handle the enormous volumes of containerized cargo, the development of EDI\footnote{Abbreviation for electronic data interchange.} in the 1960s and 1970s built the basis for the first generation of digital transformation in the maritime industry. Knowing that efficient container transportation and handling is highly dependent on the efficiency of all involved organizations and the handover of containers in between, the need for inter-organizational systems supporting a paperless communication became increasingly apparent.

One of the first EDI-based port community systems (PCS), enabling an electronic document exchange between actors involved in port operations, started in 1983 with DAKOSY\footnote{https://www.dakosy.de/en/solutions/} in the Port of Hamburg (Germany). A PCS can be defined as an inter-orga\-ni\-za\-tional system that elec\-tronically integrates heterogeneous compositions of public and private actors, technologies, systems, processes, and standards within a port community \citep{VanBaalen2009,Heilig2016}. This development of a PCS was supported by the development of the UN/EDIFACT message standards, and specific message standards for the maritime industry in the late 1980s. Important paper documents, such as the bill of lading (BoL), were transformed in the late 1980s into electronic documents. Still, the availability and quality of a PCS is seen as an essential determinant for a sustainable growth and competitiveness \citep{wiegmans2008port}. Moreover, a PCS can build the basis for establishing a \textit{single window}, ``as a facility that allows parties involved in trade and transport to lodge standardized information and documents with a single entry point to fulfill all import, export, and transit-related regulatory requirements.'' \citep{UNECE_2005} 

In the late 1980s, first commercial TOS, such as CITOS\footnote{https://www.singaporepsa.com/our-commitment/innovation} in 1988 and Navis\footnote{http://www.navis.com/timeline} in 1989, were developed and henceforth built the foundation for planning and automation in container terminals. Generally, a TOS can be defined as an information system aiding an integrated management of core terminal processes \citep{Heilig2016}. Major advances in ERP\footnote{Abbreviation for enterprise resource planning.} systems during the 1980s, driven by companies like SAP, fertilized the idea to develop TOS for improving the integration of different terminal activities. A TOS commonly integrates different sub-systems and technologies to manage and monitor the flow of cargo and handling resources, e.g., based on an integration with equipment control systems. Common TOS support EDI standards, such as UN/EDIFACT. If available, a link to the PCS is established to enable the exchange of certain information with other port-related actors over a shared platform. The integration of different internal systems and applications was essential to support individual terminal operations like berth and yard activities.  

\subsection{Second Generation (1990s -- 2000s): Automated Procedures}

The adopted information technologies and systems, such as TOS, provided an essential foundation to drastically increase the automation in container handling procedures during the 1990s and 2000s. The first automated container terminal was the ECT Delta Terminal in Maasvlakte Rotterdam (Netherlands) opened in 1993. It introduced automated guided vehicles (AGVs) and automated stacking cranes (ASC) to handle transports between the quay and container stacks, and within the container stacks, respectively. This major step towards automated container terminals required a seamless integration between the automated handling equipment and the TOS containing all work orders. The trend of using information systems as a backbone to further automate and to further increase the visibility in port operations continued during the mid and late 1990s. In particular automatic identification technologies (e.g., real-time locating systems -- RTLS) and positioning technologies (e.g., global positioning system -- GPS) were introduced in the mid 1990s to improve the efficiency and safety of port operations. Similar applications could be found in global supply chains \citep[see, e.g.,][]{leung2014aligning}. In the late 1990s, first optical character recognition (OCR) systems were launched for supporting inspection procedures. This included the installation of OCR systems in the gate area as well as image-based damage inspections, which were often combined with the capabilities of laser and video technologies, for instance, to detect container damages \citep{Heilig2016}. Also other information systems, such as vessel traffic services, used by port authorities to monitor and control vessel traffic within the port, benefited from the application of automatic identification systems in the late 1990s, allowing the tracking of vessels as a means to prevent collisions. After facing  severe traffic problems, the first information system approaches for managing truck appointments were introduced in the beginning of the 21st century. At the Los Angeles/Long Beach ports (USA), for instance, the development of the first truck appointment system started in 2002 in response to state legislation aiming to reduce truck queuing at terminal gates and to mitigate vehicle emissions \citep[see, e.g.,][]{giuliano2007reducing}. In the meantime, the development of automated container terminals proceeded apace resulting in the most modern Container Terminal Altenwerder (CTA) in the Port of Hamburg (Germany) in 2002. Furthermore, it can be observed that there was a growing interest in e-commerce systems in the late 1990s as a result of the dot-com boom, for example, to facilitate trade and shipment management between carriers, shippers, and forwarders. INTTRA\footnote{http://www.inttra.com}, developed in 2000, for example, is still the leading e-marketplace for the maritime industry providing an industry network and various functionality to support maritime shipping commerce. The global economic crisis of 2008-2009 led to a more stringent evaluation and selection of ports regarding several decision variables (e.g., cost, capacity, accessibility, connectivity, eco-friendliness) and cargo shifted between ports \citep{laxe2012maritime}. This has intensified the competition among ports drastically. According to \cite{pallis2010seaports}, a structural implication of the economic crisis was that sustainable performance can be achieved through two key strategies. While the first strategy aims to strengthen the cooperation between ports, the second strategy focuses on improving the coordination between port actors to solve, e.g., accessibility problems. Especially the current phase of digital transformation, discussed in the next subsection, aims to support these two strategies. 

\subsection{Third Generation (2010s -- Today): Smart Procedures}
\label{subsect:thirdgen}

While the first and second generations mostly focused on establishing the foundation for improved information flows in terminals and port communities as a basis for automation and  information exchange between different stakeholders in a local or global context, the on-going third generation of digital transformation aims to facilitate real-time communication to further improve the visibility, automation, coordination, collaboration, and responsiveness in intra- and inter-organizational processes in the port community and beyond. On the other hand, a purposeful integration and exploitation of available data sources shall open up new possibilities to support, improve, or adapt processes and business models. 

As described in the beginning of this section, it can be seen that current initiatives and projects in the context of \textit{smart ports} are increasingly demanding methods and solutions supporting their business analytics. With respect to container terminals, potential business analytics applications are discussed in Section~\ref{sect:dataminingterminal}. Still, a future challenge is the analysis of data in order to make netter (e.g., more efficient) decisions and to further automate intra- and inter-organizational processes as well as overall port operations including administrative procedures. A main performance indicator is their capability to pro-actively and quickly respond to changes and errors. The implementation of this vision requires multidisciplinary knowledge and is highly dependent on a successful collaboration between industry and academia. At the same time, we see that the success of those initiatives is highly dependent on the willingness of port actors to participate. While the traditionally asynchronous information exchange allowed actors to perform activities and decisions almost autonomously, new approaches may require an active and on-going information exchange and collaboration between the port and involved actors to partly contribute to a common good. Although this causes not only enthusiasm, maritime ports, especially terminal operators as main stakeholders, need to continue working on solutions for solving major issues related to the flow of cargo and logistics services in order to stay competitive. The current development and adoption of modern information systems further indicate that main actors, such as port authorities and terminal operators, increasingly extend their traditional business scope by acting as an information integrator and provider. Moreover, the impact of the digitalization may further increase security spendings for addressing resulting cybersecurity issues, especially after recent cyberattacks, such as Petya\footnote{See, e.g., \url{https://www.porttechnology.org/news/digitization\_spurs\_port\_security\_spending}.} in 2017. To summarize, the new developments will lead to a flood of complex data that need to be handled by advanced methods, tools, and information systems. In the following, the concept of business analytics is introduced as a practice for addressing current potentials and challenges related to the use of data. An in-depth analysis and discussion of the three generations of digital transformation in maritime ports is presented in \cite{heilig2017analysis}.

%

\section{Business Analytics: A Brief Introduction}
\label{sect:businessanalytics}

\begin{figure}
		\centering
		\includegraphics[width=.6\linewidth]{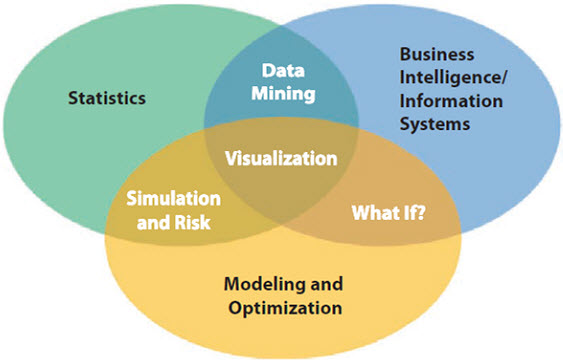}
		\caption{Topics related to business analytics \citep[see][p.\,33]{Evans_2017}
		}
		\label{fig:Evans_BA_topics}
	\end{figure}

A common definition of business analytics is the use of data, information technology, statistical analysis, quantitative methods, and mathematical or computer-based models so that managers gain improved insights about their business operations and make better, fact-based decisions. Fig.~\ref{fig:Evans_BA_topics} shows the methods, tasks, and research areas involved in modern business analytics, i.e., the integration of BI/information systems, statistics, and modeling and optimiziation. These core topics are traditional ones. The `more modern' components are shown in the intersections, and there have been a lot of improvements and influential developments with respect to methods and tools, hardware and software \citep{Evans_2017}.

\subsection{Types of Business Analytics}

\begin{figure}
		\centering
		\includegraphics[width=.8\linewidth]{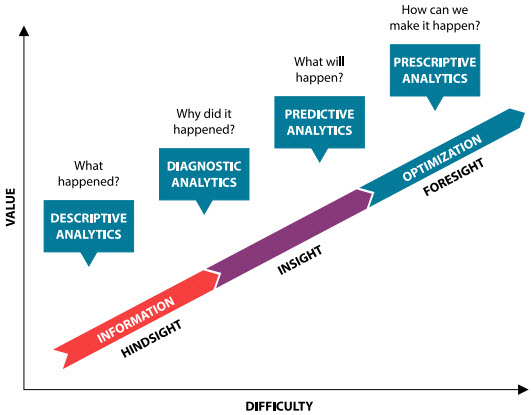}
		\caption{Types and scope of business analytics according to Gartner 
		}
		\label{fig:Gartner_value_escalator}
	\end{figure}

The concept of \textit{predictive analytics} should be viewed in relation to other types of business analytics that evolved over time. Similar to the view above but with a slightly different focus, four types of business analytics are distinguished by Gartner: descriptive, diagnostic, predictive, and prescriptive analytics. \textit{Descriptive and diagnostic analytics} can be regarded as tasks of sensing and responding, which is a more passive perspective whereas \textit{predictive and prescriptive analytics} are focused on predicting and acting \citep[see, e.g.,][]{Evans_2017,Lustig_etal_2010,Davenport_2007}. Fig.~\ref{fig:Gartner_value_escalator} shows the `analytic value escalator' with value gained from higher levels of analytics maturity (and difficulty). The value can be regarded as competitive advantage, or it can at least be turned into it. 

By performing business analytics, managers want to gain enhanced understanding of data, content, and meaning. Nowadays, the problem is not to produce or collect and save data. Data is produced more or less permanently and throughout an entire organization in internal data sources and beyond in external data sources. The problem is to unlock (hidden) value out of the enormous amount of complex data. This results in 
\begin{itemize}
\item increase of the managers' ability to make informed and better decisions, implying that decisions can be made faster without sacrificing the decision quality
\item increase of operational excellence within a company
\item better processes at the interfaces of a company by having a better understanding of customer needs as well as of suppliers' capabilities
\item establishing new business models
\end{itemize}

To summarize, the value proposition of business analytics is that it helps companies, such as terminal operators, to achieve strategic objectives \citep[see, e.g.,][]{Rios_2013}. However, this is not done automatically, i.e., it cannot be overemphasized that someone has to take action on data and results of business analytics. For example, presenting real-time data on a dashboard does not solve any real-world problem in terminal operations, but it can help to make better decisions and take better actions. Furthermore, it is common to have difficulties with the implementation of BI systems and tools for data analytics because there are typically different software sub-systems running, making the software and data integration difficult \citep[see, e.g.,][]{Rushmere_2017}.

The basic idea of data-driven business analytics is closely related to the common hierarchy of data, information, and knowledge. From a bottom-up perspective, data is the basis for information by understanding and interpreting data in a specific context. Furthermore, there are `tools' helping for a better data understanding such as visualization, aggregation, etc. Based on information, knowledge can be derived or created, e.g., by pattern detection, confirmatory data analysis, and causal interference, taking a specific application context into account. Top-down, knowledge can be used to create new information which can be encoded in data.
\textit{Data science} and \textit{data mining} are closely related, and in fact there is no widely accepted clear definition or differentiation. For example, according to \cite{Dhar_2013}, data science is regarded as the study of the generalizable extraction of knowledge from data. Currently, one differentiating factor might be the type of data, i.e., data science seems to incorporate new technologies, e.g., related to big data \citep[see, e.g.,][]{Provost_etal_2013}.

	\begin{figure}
		\centering
		\includegraphics[width=.98\linewidth]{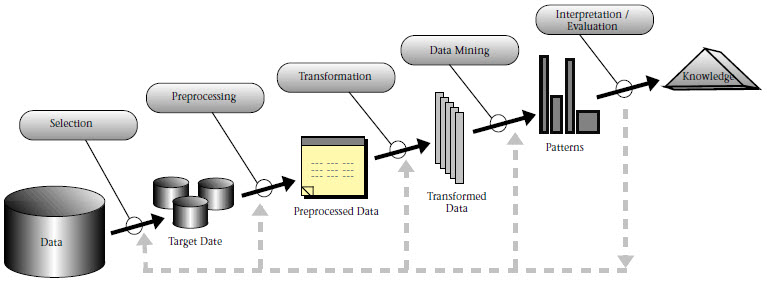}
		\caption{Steps of the KDD process  \citep[see][p.\,41]{Fayyad_etal_1996}}
		\label{fig:KDD_process_Fayyad}
	\end{figure}
	
With respect to a process view of data science, there are two well established process models. One model describes the required steps to be performed sequentially as well as in a  cycle for improving specific steps for knowledge discovery in databases (KDD process; see Fig.~\ref{fig:KDD_process_Fayyad}). Here, data mining is considered to be a part of the entire knowledge discovery process. This shows, that, e.g., recognizing patterns is not the final goal but only one step in order to derive useful knowledge. Furthermore, the preliminary steps of selecting, preprocessing, and transforming data are mentioned. These steps are often underestimated in terms of importance and workload by practitioners without deeper knowledge of KDD and data mining.

\begin{figure}
		\centering
		\includegraphics[width=.7\linewidth]{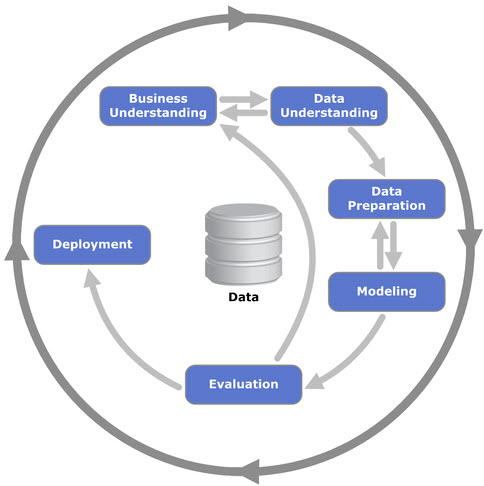}
		\caption{Six phases of CRISP \citep[see][]{Chapman_etal_1999}}
		\label{fig:CRISP}
	\end{figure}

The second model shows a \textit{cross industry standard process} for data mining (CRISP or CRISP-DM, developed by an industry consortium around 1996; see Fig.~\ref{fig:CRISP}) with six high-level phases. Similar to the KDD process, these phases are not strictly sequential but iterative, i.e., typically at a specific phase one previous phase or step has to be redesigned or changed in order to gain improvements. The reference model allows the possibility of going back and forth between different stages, or, even more strict, it is said that moving back and forth is required (although backward arrows are not shown explicitly between all phases). The shown arrows indicate 	the most important and frequent dependencies only.

The six high-level phases are still a good generic description of a general analytics process. However, details of the phases need to be updated in order to cope with current developments and problems related to big data and `modern' data science. 

\subsection{Data Mining Methods}
Data mining can be used for either discovery (of patterns among data) or for verification. Discovery can be partitioned into prediction and description. Both basic approaches aim at deriving useful information and finally knowledge out of given data. The difference is that prediction requires a target variable in the data with its value to be predicted whereas description needs no such target variable (and often problem settings simply do not have one in data). Therefore, prediction is often related to the term supervised learning while description is related to unsupervised learning.

\subsubsection{Descriptive Approaches}
Descriptive approaches can further be subdivided into segmentation (or clustering) tasks and association rule mining/sequence mining.

\textit{Clustering} is very popular for analysis of unstructured multivariate data. The aim of this unsupervised learning approach is to discover clusters among a given data set, i.e., homogeneous classes or subgroups of observations or variables. In real-world applications, the underlying assumption is that there is not one cluster only but that the heterogeneous data set can be separated into `natural' groups familiar to the domain experts. A typical application for marketers is to use consumer profiles and demographics in order to find customer groups so that campaigns can be run more effectively and efficiently. For a deeper look at  methods and algorithms see, e.g., \cite{Izenman_2008}.

The objective of \textit{association rules} is to find important  regularities in data reflected in associations. These are reflected in co-occurrence relationships among data items. A very common application of association rule mining is market basket data analysis. 
Sequence mining or sequential pattern mining also takes the sequence of purchasing into account.   
Details can be found, e.g., in \cite{Liu_2011}.

\subsubsection{Predictive Approaches}
In predictive approaches, past observations or training data are collected in a set of $n$ samples $(\mathbf{x}_i,y_i), i=1,2,...n$ that are used for estimating a function or model $f(\mathbf{x})$ ($\mathbf{x}$ is a vector of data). The training data includes the correct output values (correct in the sense that it is not a prediction but a value from the past) $y_i$, the target variable or dependent variable. This model $f(\mathbf{x})$ can be used for predicting an output value for given input values. In machine learning, this task is denoted as predictive/supervised learning. In general, machine learning is devoted to the development of algorithms to automatically extract patterns and generate a model \citep{murphy2012machine}. Prescriptive approaches can further be subdivided into classification tasks and regression tasks. 

\textit{Regression tasks} are given if a real value has to be estimated, i.e., predicted. For example, the prediction of a stock price based upon historical time series data of that stock price can be regarded as a regression task. Here, the output $y$ is real-valued, i.e., the target variable is the price, and the, e.g. daily, prediction can be a specific number. The quality of prediction is usually measured as a specific difference between the prediction and the real value.

\textit{Classification tasks} are given if an indicator function or class boundary has to be learned and estimated in order to divide samples into categories (or classes with a class label). The target variable is the class label, it is a categorical one. For a binary classification problem, the indicator function shows either 0 or 1 (or, e.g., +1 and -1), and the function separates the input space into two regions related to different classes. 


An example is the prediction of vessel arrivals related to an assumed or estimated time of arrival. The task of vessel arrivals, for instance, can be modeled as regression task or classification task: either the value of earliness/lateness is predicted, e.g., in minutes or hours, or -- more roughly but not necessarily worse, and sometimes even more appropriate with respect to the real-world task to be solved -- a class of earliness/lateness is predicted, e.g., with five classes `very early', `early', `on time', `late', `very late' and an appropriate definition of those classes. 

This example indicates that there are different ways of modeling more or less the same real-world task. The choice of the model type depends on (a) the aim of the decision maker (e.g., is a real-value prediction useful or mandatory for the planning of terminal operations or is a rough classification appropriate or even better), (b) available data, and (c) the method(s)/algorithm(s) to be used.

 Depending on the chosen model or task type, there are different well established methods or algorithms available. For an introduction to machine learning with focus on predictive learning see, e.g., \cite{Cherkassky_2013}.

\subsubsection{Prescriptive Approaches}

Prescriptive analytics is closely related to optimization approaches, aiming to identify the best alternative or alternatives regarding a minimization or maximization objective or considering multiple objectives, respectively. As shown in Fig.~\ref{fig:Gartner_value_escalator}, it is about general questions of planning, like ``how can we make it happen?'' or ``what shall we do to minimize the truck turnaround times?'' In this regard, it aims to incorporate information and knowledge, extracted through descriptive and predictive analytics, into optimization and simulation approaches, for instance, to better take into account uncertainties, such as concerning demands, arrival times, and disruptions. To put it simple, the novel term is aimed at linking the data-driven perspective with the optimization perspective. 


\section{Data Mining Applications in Container Terminals}
\label{sect:dataminingterminal}

Given that means of descriptive and predictive approaches become increasingly important in the current phase of digitalization, this section provides an overview on potential applications with respect to the main operations areas in container terminals at the quayside, yard, and landside. This includes a brief overview on academic works applying data mining methods to produce more accurate forecasts as well as to better understand and address certain problems in container terminals. Before discussing specific applications in container terminals, it should be noted that several works have proposed models to predict the container throughput, mostly based on time series analysis \citep[see, e.g.,][]{gao2016forecasting,pang2017forecasting}, or container flows between ports \citep[see, e.g.,][]{tsai2017using}. Another essential application in ports, which is also discussed as an important part of the digitalization, is the use of data analytics for improving customs and security inspections \cite[see, e.g.,][]{ruiz2017efficient,ruiz2014hybrid,Jaccard_2017}.


\subsection{Quayside Operations Area}

At the interface between seaside and landside operations, the main focus of the quayside operations area is on the discharging and loading of sea-going vessels using quay cranes (i.e., ship-to-shore gantry cranes). Moreover, it involves the horizontal transport of containers between quay wall and the yard operations area, e.g., using AGVs or straddle carriers. Besides providing modern equipment ensuring high productivity, it is important to efficiently allocate and schedule resources (e.g., berths, quay cranes, vehicles). In this regard, the quayside planning is dependent on many (external) factors, such as regarding vessel arrival times, vessel call patterns, peak demands, and the handling capacities and capabilities of the quayside equipment. Different information technologies and systems are specifically used to collect and manage operational data at the seaside, including:

\begin{itemize}
	\item \textit{Automatic identification system (AIS)}: A technology that supplements radar systems for tracking vessel positions with the primary objective of avoiding vessel collisions. After enabling the communication with satellites, referred to as S-AIS, the technology nowadays supports a real-time monitoring of vessels. AIS data messages include information about the vessel (e.g., MMSI\footnote{Abbreviation for maritime mobile service identity.}, vessel type, length, width, draught) and voyage (e.g., port of destination, speed and course over ground, heading). Using this data, several vessel tracking web services have been established (e.g., \textit{VesselFinder}, \textit{FleetMon}).
	\item \textit{Vessel traffic service (VTS)}: A VTS includes functionality to collect, analyze, and disseminate data, especially to navigate vessels in busy, confined waterways and port areas \citep{filipowicz_vessel_2004}. The information system integrates various subsystems and technologies, including AIS, vessel movement reporting systems, radar systems, radio communication systems, traffic signals, and video surveillance.
\end{itemize}

Although operations at the seaside are increasingly supported by information technologies and systems \citep[see][]{Heilig2016}, they are still affected by disruptions and uncertainties resulting from a lack of reliable information and forecasting. This includes delays and overpunctual vessel arrivals, weather conditions, tidal conditions, traffic congestion, and equipment breakdowns. With respect to quayside planning, many variations of different optimization problems have been discussed in the literature, in particular the berth allocation problem, quay crane allocation problem, and quay crane scheduling problem. Some of the discrete problem formulations consider uncertainties by using stochastic variables, for instance, stochastic arrival and handling times in the berth allocation problem \citep{bierwirth2010survey}. Stochastic programming, for instance, has been proposed as a means to address uncertainties in berth and quay crane assignments by taking into account different risk preferences of decision makers \citep[see, e.g.,][]{zhou2008study}. 

Having various sources and large amounts of operational data, the application of data mining is also attracting interest in both industry and academia. A strong research focus is on the analysis of AIS data for identifying patterns and anomalies concerning vessel operations and maritime traffic. Most related studies analyze vessel behavior patterns \citep[see, e.g.,][]{arguedas2017maritime,de2012machine} and anomalies \citep[see, e.g.,][]{lei2016framework,ristic2008statistical} or propose means to reduce the risks of vessel accidents \citep[see, e.g.,][]{hanninen2014bayesian2,zhang2015method}. Putting the focus back on container terminals, we identify only a few applications that are currently discussed in the literature with respect to quayside operations.

\subsubsection{Vessel Arrival Times}

While being important for an efficient planning of subsequent terminal operations, reliable forecasts about the actual arrival times of vessels are still scarce in many seaports. This may lead to unused terminal capacities and longer vessel waiting and turnaround times. Means to predict arrival times further allow to operate vessels more efficiently in terms of emissions. In this context, slow steaming and virtual arrival policies are currently discussed in the literature \citep[][]{meyer2012slow}, taking into account, for instance, the impact of tides. In the context of container terminals, few studies address the prediction of vessel arrival times. \cite{fancello2011prediction} propose a feedforward neural network for estimating ship arrival times in order to better determine capacity demands, for which an optimization model is used. The approach aims to reduce the number of additional workers in working shifts that need to be planned to cover uncertain demands. \cite{pani2014data} present a classification and regression trees (CART) model to further reduce the range of uncertainty of vessel arrivals using data from the Transshipment Container Terminal (TCT) of Cagliari (Italy). Compared to related works, the paper specifies in detail the steps taken in the KDD process. The authors demonstrate how the model can be used to identify the causes for delays. In another work, \cite{pani2015prediction} treat the problem as classification and assess different algorithms (logistic regression, CART, and random forest) using data from the Port of Cagliari (Italy) and the Port of Antwerp (Belgium). Besides vessel data (e.g., physical structure, previous port of call, position), the authors consider weather conditions, such as account geostrophic wind speeds, wave heights, peak wave periods, and wave directions. Using the Gini importance measure, measuring the relevance of input variables, the high impact of weather conditions on vessel arrival uncertainty is highlighted.  \cite{kim2017early} propose a modified framework of case-based reasoning (CBR) for the early detection of vessel delays using real-time S-AIS vessel tracking data in addition to historical data (e.g., data from bill of ladings). The approach allows to detect delays in real-time and predict movement patterns of a vessel until its arrival. The authors further highlight the potential improvement of predictions when using real-time data.
	
\subsubsection{Berth Operations}

To predict the performance of vessel loading and discharging operations, \cite{gomez2015development} propose a neural network that takes into account operational data (e.g., berthing time, number of containers, number of gangs, vessel beam size) as well as wind conditions (e.g., average wind speed, wind direction) during berthing of respective vessels, provided by a Spanish container terminal. By analyzing training errors, the authors highlight the important role of wind conditions. 

For addressing the berth allocation problem in bulk terminals, seeking to identify the berthing position and berthing time of bulk carriers, \cite{de2017machine} recently propose a machine learning approach for selecting optimization algorithms dependent on the scenario at hand. A k-nearest neighbors algorithm is proposed to classify each problem instance based on its features. Taking into account the historic performance of algorithms in solving similar problem instances, a ranking of algorithms is generated for each problem instance. Compared to other approaches, the study shows that data mining cannot be only used for analyzing operational data, but may also aid the selection of appropriate planning methods and tools.

%
	
\subsection{Yard Operations Area}

Yard operations mainly involve storage and stacking logistics \citep{steenken2004container} and serve as a buffer between seaside and landside operations. Several complex planning and optimization problems result from yard operations, such as yard allocation problems, post-stacking problems (e.g., remarshalling, premarshalling, and relocation problems), crane scheduling, etc. \citep[see, e.g.,][]{caserta2011container}. The performance of yard operations is constrained by several factors, including the quay wall throughput (per year / in peaks), the size and shape of the yard area, characteristics of containers (e.g. type, size, weight, destination port), the handling performance of crane systems, and handling equipment \citep{boese2011container}. These factors can have an effect on important performance indicators, such as on container dwell times (i.e., the time a container spends at the terminal), handling performance and utilization of equipment, and operational costs. Different information systems and technologies are in place to support yard operations, such as:

\begin{itemize}
	\item \textit{Terminal operating system (TOS)}: Functionality for registering new containers and tracking their position within the container yard is provided by the TOS. In particular, automated transfer cranes (ATC) rely on the availability and accuracy of job and container data from the TOS to autonomously perform yard moves.
	\item \textit{Automated transfer points for trucks}: Some container terminals have implemented automated transfer points at the yard to identify and serve incoming trucks. After following the instruction, the driver must leave the cabin and confirm the yard operations by showing a driver's card at the bay station. The latter increases safety and enables the identification of containers based on job data stored on the smart card. 
\end{itemize}

While related optimization problems have been intensively approached and discussed in recent decades, only a few works apply data mining methods to gain insights from operational data related to yard operations.

\subsubsection{Container Dwell Times}

Prolonged container dwell times result in a high storage yard occupancy and may result in adverse effects on the terminal productivity and throughput capacity. While reducing dwell times increases the yard throughput capacity, storing containers in the yard over a longer time may also result in higher revenues earned from demurrage fees. \cite{moini2012estimating} analyze different methods to predict dwell times at terminal yard operations areas at a US container terminal: na\"{i}ve Bayes (NB), decision tree (C4.5) and a hybrid Bayesian decision tree (NB tree). Using the well-performing C4.5 model, the authors further assess the impact of changes in determinants on the container dwell times, yard throughput capacity, and terminal demurrage revenues using three scenarios: changing the status of containers from empty to full, closing truck gates in low volume conditions, and changing the ocean carrier. Although more in-depth analysis is needed, the authors demonstrate the impacts of changes and trade-offs between container dwell times and demurrage fee revenues. \cite{kourounioti2016development} analyze the determinants of container dwell times by applying an artifical neural network (ANN) with backpropagation using a data set of 13733 import containers from the TOS of a container terminal in the Middle East, containing information related to the containers (e.g., arrival/departure times, size, status, type, date of customs inspection if applicable, dwell time), ocean carriers (e.g., name, assigned vessel, port of origin), and trucks (e.g., departure time from the terminal's gate). Using different sets of independent variables, the authors test their impact on the model's accuracy and show that accuracy can be improved by considering more information, whereas the measured accuracy is with 65.17\% not very high for the best case. \cite{gaete2017dwell} propose a framework of container storage assignment policies using container dwell time classes based on different classification algorithms, including NB, lazy learning (KNN; k-nearest neighbor algorithm), and rules induction learning (OneR, JRip) techniques. The authors use a data set from the Port of Arica (Chile) containing a total of 151,640 import container movements. Based on the results of the classification algorithms (JRip, KNN), a discrete event simulation model of the import processes at the Port of Arica is proposed to evaluate the impact of different stacking policies. The results demonstrate that an appropriate preprocessing and preparation of operational data, as advocated in the KDD process and CRISP (see Sect. \ref{sect:businessanalytics}), leads to a substantial reduction of re-handling activities.
	
\subsubsection{Container Stacking}	

Container stacking policies for containers have been widely discussed in the literature. Due to the ever-growing requirements to better use the space of container terminals and the impact of larger vessels, a higher yard utilization and a reduction in the number of reshuffles are desired. Besides advanced optimization and simulation approaches, only a few studies incorporate data mining methods. \cite{jin2004intelligent} present an “intelligent simulation method” based upon fuzzy ANNs for the regulation of container yard operation including the system status evaluation as well as the operation rule and stack height regulation. A two phase approach is proposed: the first phase of the regulation process forecasts the quantity of incoming containers. The second inference phase decides on the operation rule and stack height, addressed as a fuzzy multi-objective programming problem with the objective of minimizing a ship's waiting time and the operation time. A comparison between results of the proposed model and current operation in 30 days shows that the total ship waiting time is reduced from 64 h to 46 h. \cite{kang2006deriving,kang2006determination} focus on the planning of storage locations for incoming containers of uncertain weight. Oftentimes the information about the weight of a container is not accurate; when dedicated weighing procedures are not in place, the weight of containers is often underestimated or overestimated. As efficient stacking strategies are highly dependent on weight information of containers, it is important to explore means to extract information from available data sets. In this regard, the authors apply different classification algorithms to better estimate the weight group of a container, which is used in a simulated annealing algorithm to determine a good stacking strategy that reduces the number of re-handlings. However, the authors indicate that even though the overall accuracy of weight classification was improved by using the classifiers, the performance of some stacking strategies became slightly worse due to certain misclassifications. They propose to further investigate this problem by considering a cost sensitive learning for the weight classification. In the meantime, the International Maritime Organization (IMO) has amended new regulations that require a mandatory verification of the gross mass of packed containers, which may help to improve the data quality. Recently, \cite{hottung2017deep} propose a hybrid heuristic tree search integrating a deep neural network to solve the well-known container pre-marshalling problem. The neural network assists the heuristic in guiding branching and pruning. The authors show that their approach finds solutions 4\% better than state-of-the-art optimization methods using real-world sized problem instances from the literature.

\subsection{Landside Operations Area} 

Landside operations involve internal transports, truck operations, and railway operations \citep{steenken2004container}. Related horizontal transport operations rely on an efficient handover of containers at the yard or in dedicated handling areas (e.g., rail or barge terminal) and might be subject to inspections. Improving those operations leads not only to a better hinterland accessibility and inland connectivity, crucial for the competitiveness of ports \citep[see, e.g.,][]{wiegmans2008port}, but also facilitates efficient connections to auxiliary and value-added logistics areas within seaports. The increasing container volumes, peak demands, and a lack of coordination, however, lead to growing traffic and congestion at container terminals and within port areas, especially in areas located in urban environments with limited space for port expansion. As those operations highly contribute to congestion, traffic accidents, emissions, and noise, they have a great impact on the sustainable development of ports. In recent years, a large number of publications has been devoted to study and improve landside and hinterland operations, such as concerning gate/truck appointment systems \citep[see, e.g.,][]{huynh2016truck}, extended gate concepts / dry ports \citep[see, e.g.,][]{roso2010review,veenstra2012extended}, and inter-terminal transportation \citep[see, e.g.,][]{heilig2017inter,Tierney_etal_2014}. However, most of the works are conceptual or focus on (combinatorial) optimization and simulation rather than on information systems and predictive analytics. Besides, many port authorities and container terminals have greatly invested in digitalization to better manage landside and hinterland operations. Meanwhile, terminal landside operations and hinterland access are supported by various information systems and technologies.

\begin{itemize}
	\item \textit{Gate/truck appointment systems}: To better balance the workload and reduce waiting times at terminal gates, many container terminals require truck companies to pre-register containers and to book an available pickup or delivery time window. The planning of gate capacities and time windows requires a good understanding of truck arrival patterns and demand. Trucks that provide all documents in advance and arrive within the time window can therefore expect a guaranteed access to the terminal and a fast clearance process. Moreover, self-service stations have been introduced allowing the truck driver to complete missing data before arrival. Some ports penalize no-shows and late arrivals or charge a fee for day-shift or peak-hour appointments. Given existing appointment systems, several shortcomings have been reported in practice  \citep{huynh2016truck,giuliano2007reducing}, including a lack of flexibility and predictability of arrivals. While truck drivers usually meet morning appointments, keeping subsequent appointments depends on the traffic and whether the previous trips have gone as expected. In this regard, analyzing the causes of high truck turn times or late arrivals as well as the identification of late arrivals may help to reduce/avoid delays or proactively react to missed appointments, respectively. 
	
	\item \textit{Port traffic management / intelligent transportation systems (ITS)}: Some ports have implemented modern port road and traffic control systems to monitor and control traffic flows within the port area. For this purpose, different technologies, in particular sensors and actuators, are applied (e.g., laser vehicle detection systems, induction loops, etc.). The collection and analysis of traffic-related data build not only the basis to analyze motion patterns, infrastructure bottlenecks, and areas with high accident risks, but also allow to timely react to certain traffic conditions, e.g., by adapting electronic traffic signals and displaying relevant information on electronic display for traffic information and control. Cargo-related traffic data is further important to evaluate the performance of truck movements, to explore movement bottlenecks, and to determine the frequency, costs, and environmental burden of recurring events. More accurate weather data and forecasts can be used to better control the traffic and warn vehicle drivers according to certain weather conditions. Moreover, the demand for an efficient parking space management is growing. In this regard, it becomes increasingly important to make reliable predictions about the availability of parking spaces in certain areas of the port. By identifying individual motion patterns and preferences of truck drivers, context-aware recommendations can be provided. As a basis, there are already many IT-based solutions in place to support the collection, management, and dissemination of traffic-related information in ports \citep[for an overview, the interested reader is referred to][]{Heilig2016}.
	
	\item \textit{Mobile applications}: Mobile devices allow a direct interaction between actors involved in port operations and are equipped with powerful computing and sensing capabilities. Analyzing contextual data may not only help to understand the situation of individuals and to predict forthcoming events in order to provide guidance and individual recommendations (e.g., recommended travel speed to reduce emissions and to benefit a series of green traffic lights). In many ports, new mobile apps have been introduced in recent years, especially for truck drivers \citep[for an overview, the interested reader is referred to][]{Heilig2016}. 
	
	\item \textit{Rail traffic management}: Besides the truck transport, a large part of cargo movements is handled via rail transport requiring information systems to efficiently manage rail operations. An example for a corresponding information system is \textit{transPORT}, which is a new rail traffic management system of the Hamburg port railway \citep{hpa_railway_2015}. The system provides data on train locations, train movements, wagon sequences, track occupations, wagon destinations, unloading/loading schedules. In the context of synchromodality, for example, analyzing available sources of (real-time) data may be useful for predicting prices, available capacities, and the performance of alternative modalities \citep[see, e.g.,][]{van2015synchromodal}.
	
\end{itemize}

While  a growing need for data mining methods can be identified in the practical context, we currently find only a few works concerned with data mining applications. In the following, we discuss the identified works in the context of their application.

\subsubsection{Port-related Truck Traffic}

The majority of identified research works is focused on the prediction of truck-related cargo volumes in seaports. One of the first series of studies looks at cargo flows and modal split in seaports of Florida (US) in order to support strategic planning regarding the prioritizing of public funds for roadway upgrades \citep{al2000truck,al2001method,al2002use,klodzinski2003transferability,klodzinski2004methodology,sarvareddy2005evaluation}. More specifically, the authors propose backpropagation ANN models to determine relevant factors and to predict inbound and outbound heavy-truck volumes in the Port of Miami (US) and, additionally, to determine the daily modal split between inbound and outbound rail and truck cargo volumes in the Port of Jacksonville (US). In general, the proposed models in \cite{al2001method}  and \cite{al2000truck} use seaborne import and export freight data of respective ports. By considering the dwell time of containers in the container terminals, representing the lead and lag times (in days) depending on the direction of cargo, the authors were able to improve the accuracy of predictions. In \cite{al2002use}, the author applies a similar methodology for the Port of Everglades (US). As an extension, a time series model is integrated to forecast future export and import container volumes loaded/unloaded into/from container vessels, respectively. The authors do not differentiate between different sizes of containers (e.g., 20-foot and 40-foot container). In later works, the transferability of the methodology has been evaluated for additional ports in Florida, namely Port of Canaveral and Port of Tampa \citep{klodzinski2003transferability,klodzinski2004methodology}. \cite{klodzinski2004methodology} further incorporate the prediction models into simulation models in order to analyze the impact of volume variations and accidents on daily port operations. \cite{sarvareddy2005evaluation} compare the performance of the previously applied ANN with a fully recurrent neural network (FRNN), which the authors apply to better consider relationships between records (e.g., turnaround time and number of trucks) and to consider dynamic temporal behavior. The proposed FRNN model proved to be less accurate, whereas the authors do not specify how the accuracy has been measured. Generally, this early series of studies, conducted in the early 2000s, clearly indicates the dependence on field studies for collecting data about truck volumes resulting in a lack of available data. As discussed above, it is nowadays possible to collect vast amounts of data using different technologies and information systems (e.g., truck/gate appointment systems). In this regard, it would be interesting to more extensively analyze the behavior of neural networks (e.g., learning rate, overfitting, etc.) depending on different configurations (e.g., number of neurons in the hidden layers, training and transfer functions)  and sample sizes, while comparing them with other predictive methods.

In the latter sense, \cite{Xie2010} apply two kernel-based supervised machine learning methods to predict the daily truck volume at seaports, namely Gaussian processes (GP) based on a full Bayesian framework and an $\epsilon$-support vector machine ($\epsilon$-SVM), and compare them against a multilayer feedforward neural network (MLFNN). As a basis for the model development, the authors use data from Bayport Terminal and Barbours Cut Terminal (BCT) at the Port of Houston (US). Moreover, the authors evaluate the transferability of the kernel-based approaches by applying the models gained from data of one terminal to the other terminal. The authors follow the idea of \cite{al2001method} to differentiate the independent variables of import and export container volumes according to the different dwell times, but do not consider the potential impact of weekdays. Instead of considering only three days of storage \citep[as in][]{al2001method}, the models consider the previous twelve days for export containers and the next twelve days for import containers, respectively. Dependent on the direction of cargo (import or export) and container terminal, four data sets have been created covering about five months of terminal operations, resulting in twelve prediction models. Given the results of the experiments it might be possible to improve the performance of the proposed neural network model by better addressing overfitting and local minima problems. 

To identify factors that have a substantial impact on freight trip generation at the Port of Kaohsiung (Taiwan), \cite{chu2010empirical} conducts a roadside intercept survey at different facilities within the port. For evaluating the use of data mining methods, the author compares the prediction accuracy of a multiple regression model, different time series models (e.g., ARIMA, exponential smoothing model), and a backpropagation ANN. The results of MAPE, mean absolute deviation (MAD), and mean squared deviation (MSD) indicate that the ANN model has the best forecasting accuracy, followed by the regression and ARIMA model, whereas the differences are rather small. In terms of temporal effects and nonlinearity in the truck volume data, the ARIMA model and the ANN model provide a better fit. 

While most of the related works use observation data from field studies, we can also identify works that conduct nationwide surveys for identifying main characteristics of truck-trip generation \citep{holguin2002truck,chu2010empirical} or propose other methodologies and give recommendations for collecting data about container truck traffic at seaports \citep{rempel2011data}.

\subsubsection{Waiting Times and Turnaround Times}

Besides the volume of cargo, the planning of landside and hinterland operations requires reliable indicators for waiting and turnaround times, for instance for a more efficient vehicle routing. While exceeding waiting and turnaround times may greatly affect the schedule of truck drayage and hinterland operators, we can find very few studies applying data mining methods to predict them.

In the work of \cite{hill2016decision} a concept for developing a decision support system based on truck arrival rates and predicted truck gate waiting times is proposed. While the focus is primarily on the system architecture and user interfaces, the authors apply an ANN model based on actual truck waiting times from an empty container depot in Northern Germany. In the experiments, considering weekdays, daytimes, and public holidays in the set of input variables and eliminating night periods increased the model accuracy. It would be interesting to further assess the performance of alternative methods, configuration settings, for instance, by using a cross-validation and different samples sizes.

\cite{van2016benchmark} compare regression and classification models for predicting truck turnaround times using random forest and CART. For the regression analysis, the authors further use a linear regression model. Other than in related studies, the authors use a simple simulation model to generate data about terminal operations (e.g., pickups, drop-offs, time in queue, etc.). Certainly, the generated data sets can represent real terminal operations only to a certain extent since most individual factors of daily operations are not considered. Nevertheless, the work somewhat outlines a methodology for benchmarking different predictive methods using existing simulation models of container terminals.  

A recent work of \cite{wasesa2017seaport} takes a macro perspective on truck turnaround times by proposing advanced means to predict the duration of truck operations in seaports using truck  trajectory data, representing all truck movements within the port. Thereby, the work contributes to the predictive analytics development using geospatial sensor-based data. The proposed methodology involves a data preparation phase where trajectory reconstruction is first applied to understand the movement of trucks based on historical GPS positions. A geo-fencing technique is used to define the area of the seaport, which determines the trucks' arrival and departure times at the seaport and thus the duration of trucks within the seaport. The authors apply a boosting algorithm, namely the gradient boosting method, known to have a strong prediction performance and robustness against overfitting. A large telematics data set, representing five million data records from over 200 trucks operating in the Port of Rotterdam (Netherlands) over a period of 19 months, is used. Similar to \cite{hill2016decision}, the authors take into account temporal effects. Moreover, previous truck durations, truck arrivals, and truck departures serve as input variables to capture travel behavior's inertia effects \citep[see, e.g.,][]{cantillo2007modeling}. The proposed gradient boosting prediction models outperform the generalized linear models used as a benchmark. To the best of our knowledge, the study is the first in applying data mining methods to deeply study and analyze contextual data of drayage trucks based on a sufficiently large data set. As such, the work builds the basis for a promising line of research to better predict and compare the performance of seaports in handling port-related truck operations.

\subsubsection{Truck Delays}

For short-term and long-term planning, identifying the causes of inefficiencies at container terminals is at least as important as the prediction of future developments. However, the literature applying data mining methods for identifying causes and anomalies in landside and hinterland operation areas is rather scarce. 

\cite{huynh2008mining} apply three decision tree models to identify causes of abnormally high truck turn times at the BCT (US), including a chi-squared automatic interaction detector (CHAID), classification and regression tree (CART), and a decision tree (C4.5). As a data basis, the authors use transactional data from gate operations (e.g., arrival at the gate queue, terminal entry time, use of chassis, etc.) and yard operations data concerning quayside operations, drawn from the TOS, over a period of eight months. Due to the higher priorities of quayside operations, the terminal operator wanted to know, for instance, whether vessel operations pose a conflict to drayage operations. The models are formulated as binary classification problems, where the indicator function is one (1) if the truck turn time (TT) is greater than one hour, and zero (0) otherwise. By analyzing the resulting decision trees, main causes for high truck turn rates at BCT could be identified. 

\hyphenation{IMP-REQD-CHASSIS}

\begin{figure}
		\centering
		\includegraphics[width=\linewidth]{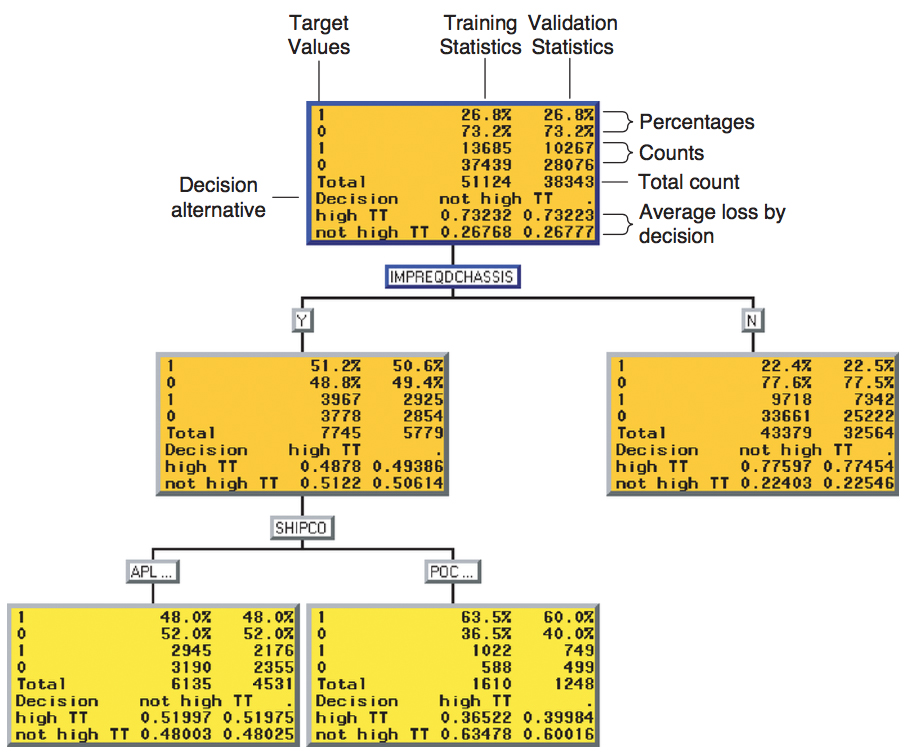}
		\caption{Example C4.5 tree \citep[see][]{huynh2008mining}. Selected input variables: IMPREQDCHASSIS (import pickup requires a chassis), SHIPCO (steamship line of the container).}
		\label{fig:turntimesdecisiontree}
	\end{figure}

An example of a decision tree of the C4.5 model is shown in Fig. \ref{fig:turntimesdecisiontree}. In this example terminal operators can easily see that the main causes for high turn times relate to the use of chassis. If an import delivery is made and it requires a chassis (\textit{IMPREQDCHASSIS}) and if the steamship line is not a chassis pool member (\textit{SHIPCO}) then transactions are likely to have high truck turn times. It can be derived that a significant delay is experienced because of the need to find and get an appropriate chassis, whereas it is even more difficult when chassis are constantly used by yard trucks at the quayside area. Other than expected by the BCT management, not the daily moves of yard cranes contributed to high truck turn times, but the lack of available chassis. Therefore, the study highlights the real benefit of data mining in identifying causes of high truck turn times at certain container terminals.

Here, the authors modelled the problem as classification task and applied decision trees for solving -- a well established approach and therefore a good choice for classification. However, at least from a scientific point of view, this case or experiment could be expanded with respect to the modeling as well as to the used algorithms. It would be interesting to see different algorithms and their pros and cons in this real-world application. For example, ANNs, support vector machines, or even random forests as a combination of decision trees could be evaluated. Furthermore, one can think of modeling this problem as a problem of association rule mining, i.e., finding associations (if-then-relationships) among coincident facts (here: usage of specific chassis and delays).

In another study, data of webcams was used to observe truck queuing patterns and to analyze the distribution of truck processing times, truck interarrival times, and truck queuing times at the entry gate of container terminals to better understand reasons of inefficient truck queuing \citep{huynh2011truck}. The authors conduct goodness-of-fit tests to identify best-fit distributions using data of two container terminals. Several implications are drawn from the distributions, such as reasons for long queues in front of the gate. For example,  long queues can be observed at the opening hour as truck drivers aim to perform as many moves per day as possible (usually they are paid per container) or in case of long turn times of other trucks within the gate.

First, at some terminals, the queues at the opening hour could be extensive because of the drayage drivers' desire to make their first move at the beginning of the day to allow for more time for subsequent moves later in the day; most drivers are paid by the move. Second, there is extensive queuing during the lunch hour at some terminals because of the policy to close for lunch. Moreover, analyzing those distributions allows to identify peak hours (e.g., arrival of a new vessel) and daily/weakly variations. The results of the authors further demonstrate that truck queuing is higher during heavy rains, thus indicating an impact of weather on terminal operations. Although some findings derived from operational data may be common sense knowledge, analyzing the data helps to accurately measure and quantify causes of inefficiencies.

\section{Conclusion and Outlook}
\label{sect:conclusion}

In recent decades, maritime ports and container terminals have invested in automation and digitalization to improve the productivity and operational efficiency of related processes. Following the developments of the current generation of digital transformation, the amount of complex data is growing at a fast pace, while remaining mostly under-processed or under-analyzed if not handled appropriately. In recent decades, quantitative research is mainly focused on optimization methods from the field of operations research. Therefore, the gap between the data, produced in and around terminal operations, and its use for terminal planning and management is growing. Business analytics represents a concept for closing this gap: to be able to use better information and knowledge in decision making processes, e.g., supported by means of optimization methods, it is essential to first process and analyze operational data. To put it concisely: a `data-driven' perspective needs to enrich the traditional `optimization' perspective.

This chapter has aimed at establishing this data-driven perspective on terminal planning and management by taking into account the current developments of the digitalization. First, the chapter has presented an overview on the three generations of digital transformation in maritime ports and then put a high level introduction to the concept of business analytics. Data mining -- as a process of discovering patterns, regularities or even irregularities in operational data -- as well as methodical approaches have been briefly explained. Given this foundation, the chapter provided a comprehensive overview on data mining applications in the context of container terminals. With respect to the different terminal operations areas, divided into the quayside, yard, and landside area, the chapter has reviewed related academic works of past decades. Most of the works focus on predictive analytics to either reduce uncertainties by data-driven forecasting models, or to better understand causes of inefficiencies or delays. In general, however, a lack of studies and applications can be identified in the field of terminal management and operations. Moreover, important methodological insights  of the data mining process, such as regarding the data preparation (e.g., data cleansing, feature selection), algorithm selection and configuration, and model evaluation, are not discussed in great detail in literature. 

Although fractional interest has been shown by a few researchers, it can be concluded that data mining research in this application domain is still in its infancy. Nevertheless, we have seen promising examples and therefore expect more research and results in the near future. Especially in terms of real-time analytics, there is a large potential to improve the responsiveness, resilience, and coordination in intra- and inter-organizational terminal operations.

%
%
%
%
%

\bibliographystyle{spbasic}

\end{document}